\documentclass[journal]{IEEEtran}

\ifCLASSINFOpdf \else \fi
\usepackage{url}
\usepackage{color}
\usepackage{color,array}
\usepackage{lipsum}
\usepackage{hhline}
\usepackage{yfonts}
\usepackage{stackengine}
\usepackage[table,xcdraw]{xcolor}
\usepackage{amssymb,amsmath,color,graphicx, amsbsy}
\usepackage{graphicx}
\usepackage{cite}
\usepackage{algorithm}
\usepackage{algorithmic}
\usepackage{amsmath}
\usepackage{amsthm}
\usepackage{amsfonts}
\usepackage{amssymb}
\usepackage{graphicx}
\usepackage{epstopdf}
\usepackage{makeidx}
\usepackage{outlines}
\usepackage[noprefix]{nomencl}
\usepackage{textcomp}
\usepackage{soul}
\usepackage{multirow}
\usepackage{bm}
\newcolumntype{P}[1]{>{\centering\arraybackslash}p{#1}}
\newcolumntype{M}[1]{>{\centering\arraybackslash}m{#1}}
\usepackage{xcolor}
\usepackage[table]{xcolor}
\usepackage{enumerate}

\usepackage{array}
\usepackage{tabularx}
\usepackage{hyperref}
\hypersetup{
	colorlinks=true,
	linkcolor=black,
	filecolor=black,
	urlcolor=black,
	anchorcolor	=black,
	citecolor=black,
}

\usepackage{graphicx}
\usepackage{amssymb}
\usepackage{amsmath}
\usepackage{longtable}

\usepackage{algorithmic}
\usepackage{array}

\ifCLASSOPTIONcompsoc
 \usepackage[caption=false,font=normalsize,labelfont=sf,textfont=sf]{subfig}
\else
 \usepackage[caption=false,font=footnotesize]{subfig}
\fi
\usepackage{url}
\usepackage{cite}

\usepackage{color,soul,graphics,graphicx,amsmath,amsfonts,amssymb,times,setspace,cite,fancyhdr,epstopdf}
\usepackage{booktabs}
\usepackage{multirow}
\usepackage{tabularx}
\usepackage[table,xcdraw]{xcolor}
\usepackage{adjustbox}
\usepackage{float}
\usepackage{longtable}
\allowdisplaybreaks

\DeclareMathSymbol{\N}{\mathbin}{AMSb}{"4E}
\DeclareMathSymbol{\Z}{\mathbin}{AMSb}{"5A}
\DeclareMathSymbol{\R}{\mathbin}{AMSb}{"52}
\DeclareMathSymbol{\Q}{\mathbin}{AMSb}{"51}
\DeclareMathSymbol{\I}{\mathbin}{AMSb}{"49}
\DeclareMathSymbol{\C}{\mathbin}{AMSb}{"43}
\DeclareMathSymbol{\Exp}{\mathbin}{AMSb}{"45}
\DeclareMathSymbol{\Prob}{\mathbin}{AMSb}{"50}
\usepackage{comment}

\ifCLASSINFOpdf

\renewenvironment{longtable}{\begin{center}\begin{tabular}}{\end{tabular}\end{center}}

\renewcommand{\toprule}[2]{\hline}
\renewcommand{\midrule}[2]{\hline}
\renewcommand{\bottomrule}[2]{\hline}

\ifCLASSOPTIONcompsoc
 \usepackage[caption=false,font=normalsize,labelfont=sf,textfont=sf]{subfig}
\else
 \usepackage[caption=false,font=footnotesize]{subfig}
\fi

\begin{document}
\title{Risk-Averse Resilient Operation of Electricity Grid 
Under the Risk of Wildfire}

\author{Muhammad Waseem,~\IEEEmembership{Student Member,~IEEE},
        Arash F. Soofi,~\IEEEmembership{Student Member,~IEEE}, and
        Saeed D. Manshadi,~\IEEEmembership{Member,~IEEE}
        \thanks{The authors are with the Department of Electrical and Computer Engineering, San Diego State University, San Dieg, CA, 92182, USA email:(mhuwaseem@gmail.com; afarokhisoofi@sdsu.edu; smanshadi@sdsu.edu)}

}

\markboth{}%
{Shell \MakeLowercase{\textit{et al.}}: Bare Demo of IEEEtran.cls for IEEE Journals}


\maketitle
{
\begin{abstract}
Wildfires and other extreme weather conditions due to climate change are stressing the aging electrical infrastructure. Power utilities have implemented \emph{public safety power shutoffs} as a method to mitigate the risk of wildfire by proactively de-energizing some power lines, which leaves customers without power. 
System operators have to make a compromise between de-energizing of power lines to avoid the wildfire risk and energizing those lines to serve the demand. In this work, with a quantified wildfire ignition risk of each line, a resilient operation problem is presented in power systems with a high penetration level of renewable generation resources. A two-stage robust optimization problem is formulated and solved using column-and-constraint generation algorithm to find improved balance between the de-energization of power lines and the customers served. Different penetration levels of renewable generation to mitigate the impact of extreme fire hazard situations on the energization of customers is assessed. The validity of the presented robust optimization algorithm is demonstrated on various test cases.
\bstctlcite{IEEEexample:BSTcontrol}
\end{abstract}
}
\begin{IEEEkeywords}
Fire hazard, extreme weather, uncertainty of renewable generation resources, risk mitigation, two-stage robust optimization, column-and-constraint generation algorithm.
\end{IEEEkeywords}

\IEEEpeerreviewmaketitle
\vspace{-.25cm}
\subsection*{Indices}
\vspace{-0.3cm}
\begin{longtable}{l p{2.7in}} 
\textit{g} $\in$ \(\mathcal{G}\) & Index of generation unit in the set of generation units\\
\textit{i} $\in$ \(\mathcal{I}\) & Index of bus in the set of buses\\
\textit{l} $\in$ \(\mathcal{L}\) & Index of line in the set of lines\\
$\mathcal{L}^{fr}_{i} (\mathcal{L}^{to}_{i})$ & Set of lines originated from (destined to) bus $i$\\
\textit{s} $\in$ \(\mathcal{S}\) & Index of solar in the set of solar generations\\
\textit{seg}   $\in$ $\mathcal{S}_g$ & Index for segment of generating unit $g$\\
\textit{t} $\in$ \(\mathcal{T}\) & Index of time in the set of time
\end{longtable}
\vspace{-0.2cm}
\subsection*{Variables}
\vspace{-0.2cm}
\begin{longtable}{ l p{2.6in} } 
\textit{\(I_t^l\)} & Binary variable representing status of line $l$ at time $t$, energized if (1); otherwise (0) \\
\textit{\(P_t^g\)} & Power dispatch of generation unit $g$ at time $t$ \\
\textit{\(P_t^{g,seg}\)} & Power dispatch of segment $seg$ of generating unit $g$ at time $t$ \\
\textit{\(P_{s,t}^{i}\)} & Real power dispatch of solar $s$ at bus $i$ and time $t$ \\
\textit{\(P_{t}^{d,i}\)} & Demand served at bus $i$ and time $t$\\
\textit{\(P_t^{l}\)} & Power flow of line $l$ at time $t$\\
\textit{\(u_{t}^{D,i}\), \(v_{t}^{D,i}\)} & Uncertainty of the demand connected to bus $i$ at time $t$
\end{longtable}
\begin{longtable}{ l p{2.7in} } 
\textit{\(u_{t}^{s,i}\), \(v_{t}^{s,i}\)} & Uncertainty of the solar generation unit connected to bus $i$ at time $t$\\
\textit{\(\theta_t^i\)} & Voltage angle of bus $i$ at time $t$
\end{longtable}
\vspace{-0.2cm}
\subsection*{Parameters}
\begin{longtable}{ l p{2.7in} }
\textit{\(c_g^{seg}\)} & Generation cost of segment $seg$ of unit $g$ \\
\textit{E} & Budget of uncertainty\\
\textit{\( K, M\)} & $K$ is a penalty factor for not serving load and $M$ is an arbitrary large constant\\\
\textit{\(\overline{P}_{t}^{i}\)} & Solar availability at bus $i$ during time $t$\\
\textit{\(\overline{P}^{g,seg}\)} & Maximum generation of segment $seg$ of generating unit $g$\\ 
\textit{\(P_{t}^{D,i}\)} & Scheduled consumer demand at bus $i$ and time $t$\\
\textit{\(\overline{P}_g (\underline{P}_g)\)} & Maximum (minimum) real power dispatch of generation unit $g$ \\
\textit{\(\overline{P}_l\)} & Maximum real power flow of line $l$\\
\textit{\(SC_t^l\)} & Fire ignition score of energized line $l$ at time $t$\\
\textit{\(x_l\)} & Reactance of line $l$\\
\textit{\(\mathcal{\triangle}P_{t}^{d,i}\)} & Deviation in demand $d$ at bus $i$ and time $t$\\
\textit{\(\mathcal{\triangle}\alpha_{s,t}^{i}\)} & Deviation in solar generation of solar $s$ at bus $i$ and time $t$\\
\textit{\(\mathcal{E}\)} & Level of risk tolerance \\
\textit{\(\psi_t^l\)} & Fire ignition score of line $l$ at time $t$
\end{longtable} 
\section{Introduction}
\subsection{Motivation}
\lettrine{R}{esilient} electricity grid reduces the vulnerability to severe fire hazard weather conditions. The average cost of power outages resulting from severe weather is \$ 18-33 billion per year. The costs can be much higher if the weather conditions are extremely fire hazard as Hurricane Ike cost the economy \$ 40-75 billion in 2008 \cite{house2013economic}. California Camp Fire in 2018, which was triggered due to a power line, killed $84$ people and resulted in \$ 9.3 billion damage to the residential property \cite{camp_fire_ref}. During $2017$ and $2018$ California fire seasons, the liable utility Pacific Gas \& Electric (PG\&E) company filed for bankruptcy \cite{pge_fire} and acknowledged unintentional manslaughter charges \cite{pge_man}. During the $2017$ fire season in the Wine Country of California, the equipment of PG\&E started numerous fires. Climate change is expected to continue and will increase the extreme weather conditions as stated by the national climate assessment. One of the ways in which power system infrastructure can cause fire ignition is the contact between conductors and vegetation \cite{jazebi2019review}. The contact occurs when the ground clearance is reduced by the conductors sag \cite{jazebi2019review_1}. During windy conditions, the wildfire spreads faster and is difficult to control, and results in increase of wildfire ignition by power lines. It is critical to have a reliable electricity grid under such scenarios. 
\par
In $2012$, California public utilities commission directed that California public utilities code sections $451$ and $399.2$(a) permit electric utilities to shut off power to customers in order to ensure public safety \cite{psps}. In Public Safety Power Shutoffs (PSPSs), power lines are de-energized during high-risk weather conditions to prevent the risk of fire ignition. Utility companies in California including PG\&E, San Diego Gas and Electric, PacifiCorp, and Southern California Edison conducted $51$ PSPS from $2013$ to $2020$, and it affected $3.2$ million consumers \cite{murphy2021}. PG\&E conducted PSPS in October $2019$, it impacted $1.8$ million consumers, and the outage lasted for more than $5$ days \cite{cnbc2019}. It is forecasted that PSPS could cost California in billions of dollars \cite{cnbc2019} in addition to the causalities due to power shutoffs \cite{fox2019}. It can be observed that PSPS is a way to mitigate the risk of wildfire ignition but it is not optimal because it de-energizes all lines in a particular region. 
\par 
In the literature, wildfire risk hazard has not been included as a part of operation planning problem. In our previous work \cite{waseem2021resilient, waseem}, a surrogate model to quantify the risk of fire ignition in power system network under severe weather conditions is presented. In continuation to that work, the objective of this paper is to operate the electricity grid under fire hazard risk, and uncertainty in demand and solar energy generation. This considers the energization of power lines to serve the load demand and prevent the fire hazard simultaneously. This paper seeks the answers to the following questions. \emph{1) Given a certain level of risk, what would be the most critical lines to energize during fire hazard severe weather conditions? 2) What is the impact of demand and solar generation uncertainties during fire hazard weather? 
3) How the distributed solar generation and its different penetration levels in the network can help to make the system resilient?}

\subsection{Literature Review}

A review of electricity grid vulnerabilities due to natural disasters is given in \cite{waseem2020electricity}. Measures to reduce fire ignition probability includes aggressive management of vegetation, higher frequency of inspections, and modification in protection equipment to limit the fault currents \cite{sce, pge_plan_report}. An optimal hardening plan to ensure resiliency and optimal functionality of the power network under extreme weather events using a tri-stage stochastic mixed-integer resilience-based design is presented in \cite{ma2018resilience}. A two-stage stochastic problem in which the first stage optimizes the post-disaster repair and the second stage optimizes the sections to harden for an efficient post-disaster recovery operation is presented in \cite{tan2018distribution}. A tri-level optimization problem to minimize the investment of hardening and estimated load shedding cost under a severe weather environment is presented in \cite{ma2016resilience}. It considers the vegetation management as a part of tri-level optimization. A resilient boosting approach using defensive islanding to prevent the cascading failures during extreme weather is presented in \cite{panteli2016boosting}. An approach based on robust optimization for enhancing the grid resilience is presented in \cite{yuan2016robust}. The resilient planning enhancement against extreme weather using four stages including resilient planning, preventive actions, corrective actions, and restoration stages is considered in \cite{huang2017integration}. The security assessment of power systems under uncertainties is examined in \cite{ciapessoni2016probabilistic}. The pre-disaster costs due to resources allocation and generation units operation are minimized based on the damage probabilities of power system components in \cite{arab2015stochastic}. The microgrids are important in resilience enhancement of the power system. The sectionalization of the distribution system into the microgrids is performed in \cite{wang2015self} to supply power to maximum isolated loads during a fault.

A proactive schedule of distributed generation units and storage devices to mitigate the impact of progressing wildfire on a distribution network is presented in \cite{trakas2017optimal}. The time-varying possibility distribution function has been proposed to enhance the resilience of day-ahead unit commitment of power systems in \cite{trakas2019resilience}. The thermal capacity of the power lines decreases due to nearby fire and it affects the operational limits. A probabilistic approach for planning the reserves optimization and generation capacities based on the stochasticity of line and generation unit failure is presented in \cite{fernandez2016probabilistic}. A unit commitment approach for improving the resilience of power networks by decreasing loading on lines that have the potential to be affected by weather events is proposed in \cite{wang2018resilience}. The remaining lines in the network are equally loaded. The decrease in line failure decreases the cascading events and load shedding.

The existing literature \cite{ma2016resilience, panteli2016boosting, yuan2016robust} and \cite{panteli2015modeling, hello2016, panteli2017metrics} present measures to enhance the resiliency of the power system under natural disasters other than wildfire. An optimal approach for the operation of the power system against progressive wildfire is presented in \cite{mohagheghi2015optimal}. This work considers the resilient operation of power system during the wildfire. People worked in during or post-wildfire scenarios but not much work has been done for the preventive measures. In the literature, quantification of conductor clashing score is not considered. In this work, the quantified risk of wildfire ignition score for each line is considered. Our work is preventing the wildfire from happening in the first place. This paper presents a two-stage robust optimization approach to ensure the resilient operation of electricity grid under extreme wildfire hazard weather conditions based on the quantified risk of fire ignition score for each power line.
\subsection{Summary of Contributions}
 The contributions of this paper are outlined as follows:
 \begin{enumerate}
 \item 
 The risk of operation of power lines is quantified to indicate the most vulnerable lines and the lines upon which the operator can rely for power transmission. A two-stage robust optimization problem for risk averse resilient operation of power system under uncertainty is formulated.
 \item The impact of deviation in demand and solar generation from nominal values on the operation cost and the resilient operation of the network under the risk of wildfire are assessed.
 \item Different penetration levels of solar generation units are considered to identify the most resilient penetration level against the risk of wildfire. The impact of distributing the solar generation on operation cost of the network under the risk of wildfire is considered.
\end{enumerate}
The remainder of the paper is organized is as follows. Section \ref{pblm_form} describes the problem formulation, while section \ref{solution_method} presents the two-stage robust optimization solution using Column-and-Constraint Generation Algorithm (C$\&$CGA). Section \ref{case_study} describes the case study and provides numerical results, while section \ref{conclusions} summarizes and concludes the paper.

\section{Problem Formulation}\label{pblm_form}
In this section, the mathematical formulation for the resilient operation of the electricity grid under the risk of wildfire ignition with considering the uncertainty in demand and solar generation is presented in \eqref{pblm_formulation}. It is formulated as a max-min problem in which the operation cost under the risk of wildfire is minimized and the impact of uncertainties on the operation costs is maximized. The objective is to minimize the operation cost and load-shedding in the network by respecting all physical and operational constraints as given in (\ref{pblm_1a}). 
The variable after each colon in \eqref{pblm_1b}-\eqref{pblm_1i} are the dual variables corresponding to each constraint. The total dispatch of generation unit $g$ is equal to the summation of the dispatch of each  segment of the generation unit as shown in \eqref{pblm_1b}. The real power dispatch of segments of each generation unit is limited in \eqref{pblm_1c}. The nodal balance equation is presented in \eqref{pblm_1d}. The physical limits of real power generation for each generation unit is presented in \eqref{pblm_1e}. The solar power dispatch and served demand are constrained in \eqref{pblm_1e_f} and \eqref{pblm_1f}, respectively. The power flow limits of lines are given in \eqref{pblm_1g} and \eqref{pblm_1h}. The quantified fire ignition score of power lines is limited by the risk tolerance as given in \eqref{pblm_1i_1} and \eqref{pblm_1i}. The uncertainties in net demand and solar generation are within a range as presented in \eqref{pblm_1l} and \eqref{pblm_1m}, respectively. The $0$ superscript in scheduled consumer demand $(P_{t}^{D,i,0})$ and available solar power $(P_{s,t}^{i,0})$ denotes the nominal value. Thus, the uncertain variables are determined using the nominal values and the uncertainty interval.
\begin{subequations}\label{pblm_formulation}
\begin{alignat}{2}
&\underset{{P_t^{D,i},\overline{P}_{t}^{i}}}{\textbf{max}} \underset{{P_{t}^g,P_t^{d,i}}}{\textbf{min}} \sum\limits_t\begin{Bmatrix}\sum\limits_{seg}\sum\limits_g c_g^{seg} P_t^{g, seg} \\
& \hspace{-2.8cm}+\sum\limits_i K (P_t^{D,i}-P_t^{d,i})\end{Bmatrix} \label{pblm_1a}\\
& \textbf{subject to:} \nonumber\\
& \sum\limits_{seg} P_t^{g, seg} = P_t^g \hspace{0.4cm} \forall seg \in \mathcal{S}_g, g \in \mathcal{G}, t \in \mathcal{T} \hspace{0.7cm} :\lambda1_t^g\label{pblm_1b}\\
& 0 \leq P_t^{g, seg} \leq \overline{P}^{g,seg} \hspace{0.4cm} \forall g \in \mathcal{G}, seg \in \mathcal{S}_g, t \in \mathcal{T}  \nonumber\\
&\hspace{5.3cm} :\underline{\mu1}_t^{g,seg}, \overline{\mu1}_t^{g,seg}\label{pblm_1c}\\
& \sum\limits_{g\in G_i} P_{t}^g+\sum\limits_{s\in S_i} P_{s,t}^{i} + \sum\limits_{l\in \mathcal{L}^{to}_{i}} P_{t}^{l}= \sum\limits_{l\in \mathcal{L}^{fr}_{i}} P_{t}^{l} + P_{t}^{d,i} \nonumber\\
&\hspace{4.5cm}\forall i \in \mathcal{I},  t \in \mathcal{T} \hspace{0.6cm}:\lambda2_{t}^i\label{pblm_1d}\\
& \underline{P}_g \leq P_{t}^g \leq \overline{P}_g  \hspace{1.4cm} \forall g \in \mathcal{G}, t \in \mathcal{T} \hspace{0.7cm} :\underline{\mu2}_t^g, \overline{\mu2}_t^g\label{pblm_1e}\\
& 0 \leq P_{s,t}^{i} \leq \overline{P}_t^{i}  \hspace{0.6cm} \forall i \in \mathcal{I}, \forall s \in \mathcal{S}, t \in \mathcal{T} \hspace{0.2cm} :\underline{\mu3}_{s,t}^i, \overline{\mu3}_{s,t}^i\label{pblm_1e_f}\\
& 0 \leq P_t^{d,i} \leq P_t^{D,i} \hspace{0.9cm} \forall i \in \mathcal{I},  t \in \mathcal{T} \hspace{0.6cm} :\underline{\mu4}_t^{d,i}, \overline{\mu4}_t^{d,i}\label{pblm_1f}\\
& -M(1-I_t^l)+P_t^{l} \leq \frac{\sum\limits_{i\in L_t}\theta_t^i - \sum\limits_{j\in L_f}\theta_t^j}{x_l} \leq P_t^{l}+M(1-I_t^l) \nonumber\\
& \hspace{3.6cm}  \hspace{0.4cm} \forall l \in \mathcal{L}, t \in \mathcal{T} \hspace{0.4cm} :\underline{\mu5}_t^l, \overline{\mu5}_t^l \label{pblm_1g}\\
& -\overline{P}_l \times I_t^l \leq P_t^{l} \leq \overline{P}_l \times I_t^l \hspace{0.4cm} \forall l \in \mathcal{L}, t \in \mathcal{T} \hspace{0.2cm} :\underline{\mu6}_t^l, \overline{\mu6}_t^l \label{pblm_1h}\\
& SC_t^l \geq {\psi}_t^l \times I_t^l \hspace{2.4cm} \forall l \in \mathcal{L}, t \in \mathcal{T}  \hspace{0.6cm} :\underline{\mu7}_{t}^{l}\label{pblm_1i_1}\\
& \sum\limits_l SC_t^l \leq \mathcal{E} \hspace{3.5cm} \forall t \in \mathcal{T} \hspace{0.6cm} :\mu8_{t} \label{pblm_1i}\\
& P_t^{D,i} \in \begin{bmatrix}P_t^{D,i,0}-\bigtriangleup P_t^{D,i},\hspace{0.1cm} P_t^{D,i,0}+\bigtriangleup P_t^{D,i} \end{bmatrix} \label{pblm_1l}\\
& \overline{P}_t^{i} \in \begin{bmatrix}P_{s,t}^{i,0}-\bigtriangleup \overline{P}_{s,t}^{i},\hspace{0.1cm}P_{s,t}^{i,0}+\bigtriangleup \overline{P}_{s,t}^{i} \end{bmatrix} \label{pblm_1m}
\end{alignat}
\end{subequations}

\begin{figure}[bht] 
    \centering
       \includegraphics[width=0.48\textwidth]{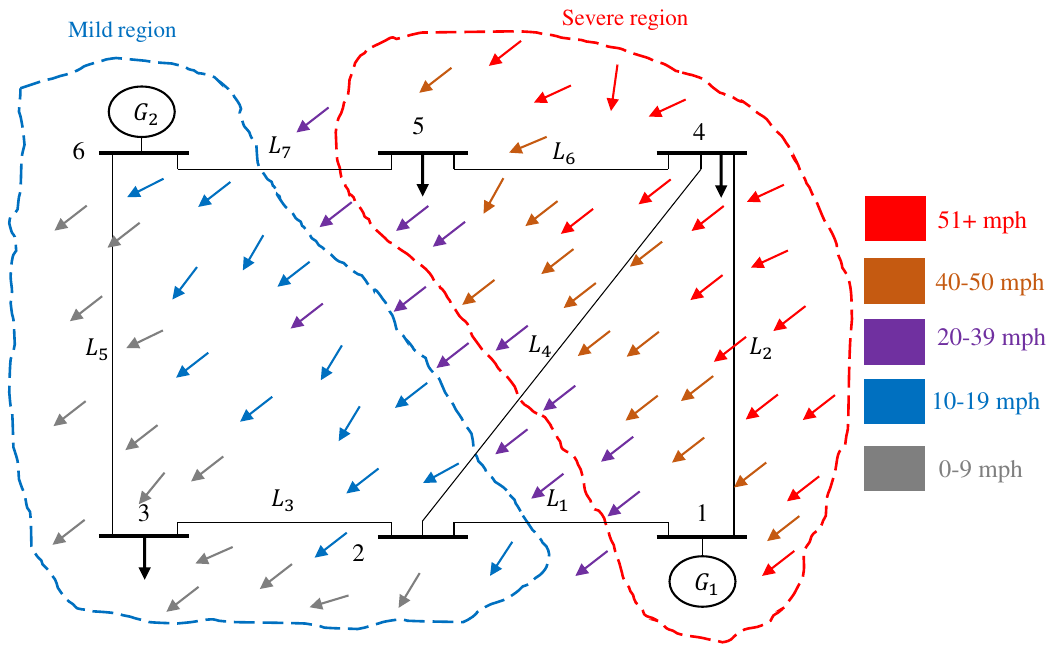}
    \caption{Different wind speeds leading to different fire ignition scores in a 6-bus test case}
    \label{lost_wind} 
\end{figure}

The quantified fire ignition score of energized power lines is obtained from our previous work \cite{waseem}. A demonstration of different wind speeds in a 6-bus test case is shown in Fig. \ref{lost_wind} where the variation in wind speed at different regions is depicted by different arrow colors as given in the legend. Different wind speeds lead to different fire ignition scores. 
The fire ignition score is determined by physically modeling the 3D non-linear vibrational motion of power lines based on Hamilton's principle. The 3D equations are simplified to 2D equations based on modeling assumptions and boundary conditions. The resulting 2D continuous partial differential equations (PDEs) are transformed into discrete PDEs using Galerkin method. These discrete equations describe both in-plane and out-of-plane non-linear vibrational motion of power lines. Runge-Kutta method is applied to solve the resultant equations. Various physical, structural, and meteorological parameters including span of line, conductor diameter, phase clearance, wind speed, wind gust, and wind direction are incorporated into the model. The fire risk hazard score is calculated based on how many points of the conductors are coming in contact by varying different features. A surrogate model using machine learning algorithm is developed to forecast the fire ignition score $\psi_t^l$. A sample of fire ignition score predictions is shown in Table \ref{samples}. 

The problem presented in \eqref{pblm_formulation} is a two-stage robust optimization problem in which the first stage (\emph{here-and-now}) decision variables minimize the operation cost of the network and the second stage (\emph{wait-and-see}) decision variables determine the realizations of worst case uncertainties. The load shedding is penalized by the penalty factor in the first stage. The second stage realizes the uncertainties in load demand and solar generation. The objective is to minimize the operation cost based on the budget of uncertainty that accounts for the uncertainty period and the conservativeness of the operator. 

{\setcounter{table}{0}
\begin{table}[h!]
    \vspace{-0.31cm}
	\small \centering
	\caption{\color{black}A Sample from Dataset in which Fire Ignition Scores are Predicted by Varying Different Features} 
	\vspace{-0.2cm}
	\begin{tabular}{M{0.4cm}M{1.27cm}M{0.49cm}M{0.70cm} M{1.29cm}M{1.21cm}M{0.54cm}} \hline\hline
	    \textbf{Span} \textbf{(ft)}  & \textbf{Conductor diameter} \textbf{(mm)} & \textbf{Wind speed} \textbf{(m/s)} & \textbf{Wind gust} \textbf{(m/s)} & \textbf{Phase clearance (ft)} & \textbf{Wind direction (\textdegree)} & \textbf{Score} [0,1)  \\ \hline 
		600& 33.03& 10& 12& 0.5&45  &0  \\
	    800& 34.02& 18& 16&0.5& 180   &0.01\\
		500& 33.03& 18& 20&0.7& 45 &  0.08\\
		1000& 31.05& 28& 30&0.5& 90  &0.12 \\
		400& 33.03& 26& 14&0.9&315 & 0.17 \\
		300& 33.03& 30& 30&0.5& 315   &0.45\\\hline\hline
	\end{tabular}
	\label{samples}
\end{table}}

\vspace{-0.4cm}
\section{Solution Method}\label{solution_method}
\begin{figure}[bht] 
    \centering
       \includegraphics[width=0.45\textwidth]{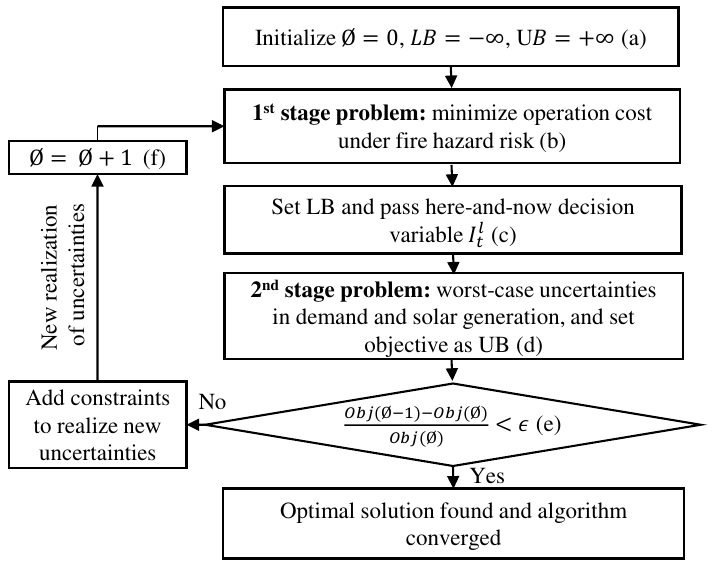}
    \caption{Flowchart of Column-and-Constraint Generation Algorithm}
    \label{ccg_algo_flow} 
\end{figure}

In this section, C$\&$CGA to solve the two-stage robust optimization problem \eqref{pblm_formulation} is described.
The general max-min robust optimization problem is given in \eqref{ccg_algo}. The objective function is given in \eqref{ccg_1} where $x$ is the first stage decision variable while $u, y$ are the second stage decision variables. The constraint with first stage binary decision variable is represented in \eqref{ro_2}. The dispatch of generation unit decision variable $y$ and realization of uncertainty $u$ in solar generation and load demand are enforced in \eqref{ro_3}. The feasible domain of first and second stage decision variables is given in \eqref{ro_4}.
\begin{subequations}\label{ccg_algo}
\begin{alignat}{2}
& \underset{{x}}{\textbf{min}}\hspace{0.1cm} c^Tx + \underset{{u}}{\textbf{max}}\hspace{0.1cm}\underset{{y}}{\textbf{min}}\hspace{0.1cm} b^Ty \label{ccg_1} \\
& \hspace{-0.5cm} \textbf{subject to:} \hspace{0.2cm} Ax \geq d \label{ro_2} \\
& Fx + Gy \geq h - Hu \label{ro_3} \\
& x \in \Omega_x, \hspace{0.1cm}y \in \Omega_y \label{ro_4}
\end{alignat}
\end{subequations}
The following are the six steps to solve the two-stage robust optimization problem.
\begin{enumerate}[(a)]
    \item Set iteration $\varnothing=0$, lower bound ($\text{LB})=-\infty$, and upper bound ($\text{UB})=+\infty$. Set budget of uncertainty $E$ and risk tolerance level $\epsilon$. Determine the realization of uncertainty $u^{*(0)}$.
    \item Solve the first stage problem
    \begin{subequations}\label{fs}
    \begin{alignat}{2}
    & \underset{{x, y}}{\textbf{min}}\hspace{0.1cm} c^Tx + e \label{fs_1} \\
    &\hspace{-0.5cm} \textbf{subject to:} \hspace{0.3cm}
     e \geq b^Ty^{(\varnothing)} \label{fs_2}\\
    & \text{A}x \geq d \label{fs_3} \\
    & \text{F}x + \text{G}y^{(\varnothing)} \geq \text{h} - \text{H}u^{*(\varnothing)} \label{fs_4} \\
    & x \in \Omega_x, \hspace{0.1cm}y \in \Omega_y \label{fs_5}
    \end{alignat}
    \end{subequations}
    \item Set $\text{LB}=c^Tx^*+e^*$ where $x^*$ and $e^*$ are the solutions obtained from (\ref{fs_1})-(\ref{fs_5}).
    \item Solve the second stage problem to find the new realizations of uncertainties.
    \begin{subequations}\label{ss}
    \begin{alignat}{2}
    & \underset{{u}}{\textbf{max}}\hspace{0.1cm}\underset{{y}}{\textbf{min}}\hspace{0.1cm} b^Ty\label{ss_1} \\
    & \hspace{-0.5cm}\textbf{subject to:} \hspace{0.3cm} \text{F}x^* + \text{G}y \geq \text{h} - \text{H}u^{(\varnothing+1)} \label{ss_2} \\
    & y \in \Omega_y, \hspace{0.1cm} u^{(\varnothing+1)} \in \text{U} \label{ss_3}
    \end{alignat}
    \end{subequations}
    \item Set $\text{UB}=c^Tx^*+b^Ty^*$ where $y^*$ is the solution obtained from (\ref{ss_1})-(\ref{ss_3}) and check if $\frac{\text{UB}-\text{LB}}{\text{UB}} \leq \epsilon$. Terminate the process if the condition is satisfied and the algorithm is converged. Otherwise, new realization of uncertainties is accomplished by adding (\ref{ab_1})-(\ref{ab_2}) to (\ref{fs_1})-(\ref{fs_5}).
    \item Set $\varnothing=\varnothing+1$ and return to step (b).
    \begin{subequations}\label{ab}
    \begin{alignat}{2}
    &  \text{F}x + \text{G}y^{(\varnothing+1)} \geq \text{h} - \text{H}u^{*{(\varnothing+1)}} \label{ab_1} \\
    & e \geq b^Ty^{(\varnothing+1)} \label{ab_2}
    \end{alignat}
    \end{subequations}
\end{enumerate}
The first stage objective renders the minimum operation cost based on limited subset of constraints while the second stage objective renders the worst-case operation cost based on worst realization of uncertainties. The second stage objective is calculated by taking the \emph{here-and-now} decision variable from the first stage as input. 
The algorithm iterates between the first and second stage problems until convergence occurs based on the operator-defined risk tolerance. The flow chart of C$\&$CGA is given in Fig. \ref{ccg_algo_flow}. If the first stage decision variable is not robustly feasible, the second stage problem only returns the worst-case uncertainty scenario $u^{(\psi+1)}$ and the dispatch is not possible. On the other hand, when the first stage decision variable is robustly feasible, the second stage problem returns the worst-case operation cost and the respective worst-case uncertainty scenario $u^{(\psi+1)}$.

The uncertainty in demand and solar generation dispatch are taken into account and are limited by budget of uncertainty. Budget of uncertainty limits the binary variables that render the worst case realization of uncertain variables. It represents the conservativeness of the system operator and the size of uncertainty period. The uncertainty variables found in the second stage comprise of polyhedral uncertainty set and the aim is to minimize the total cost of first and second stages based on budget of uncertainty. Different values of budget of uncertainty ($0,1,5,10,20$, and $50$) are considered in the case study and the impact on operation cost is evaluated. It is considered as a relaxation to the primary problem and chosen by the system operator. 

The sub-problem in \eqref{ss_1}-\eqref{ss_3} is re-formulated as a dual problem since the worst-case scenarios occur within the polyhedral uncertainty set.
The dual formulation of problem \eqref{pblm_formulation} is given in \eqref{dual}. The objective of dual form is given in \eqref{dual_1a} and the dual constraints are given in \eqref{dual_1b}-\eqref{dual_1g_E}. Each dual constraint corresponds to a primal variable as indicated in parentheses.
\begin{subequations}\label{dual}
\begin{alignat}{2}
&\textbf{max} \sum\limits_t\begin{Bmatrix}\hspace{-0.1cm}\sum\limits_g\sum\limits_{seg} -\overline{P}^{g,seg}\overline{\mu1}_t^{g,seg}+\sum\limits_g(\underline{P}_g\underline{\mu2}_t^{g}-\overline{P}_g\overline{\mu2}_t^{g})\\
& \hspace{-7.2cm}- \sum\limits_i \{P_t^{D,i} \overline{\mu4}_t^{d,i}+ \triangle P_t^{d,i} P_t^{D,i}\overline{\mu4}_t^{d,i} u_t^{D,i}\} \\
&\hspace{-7.7cm}- \sum\limits_l \{\overline{P}_l (I_t^l\underline{\mu6}_{t}^{l}+I_t^l\overline{\mu6}_{t}^{l})+M(1-I_t^l)\\
& \hspace{-7.80cm}(\underline{\mu5}_{t}^{l}+\overline{\mu5}_{t}^{l})\} - \mathcal{E} \mu8_{t}+ \sum\limits_{l}\psi_t^l  I_t^l\underline{\mu7}_{t}^{l}\\
& \hspace{-5.1cm}-\sum\limits_{s}\overline{P}_{t}^{i} (\overline{\mu3}_{s,t}^{i}-\triangle\alpha_{s,t}^{i}\overline{\mu3}_{s,t}^{i}u_t^{s,i})\end{Bmatrix} \label{dual_1a} \\
& \textbf{subject to:} \nonumber\\
& \hspace{0.2cm} -\lambda1_t^g + \sum\limits_{i\in B_{g}}\lambda2_{t}^i + \underline{\mu2}_t^g - \overline{\mu2}_t^g=0 \hspace{0.5cm}\forall g, \forall t \hspace{0.4cm}(P_t^g)\label{dual_1b} \\
& \sum\limits_{i\in B_{s}}\lambda2_{t}^i- \overline{\mu3}_{s,t}^i \leq 0 \hspace{2.4cm}\forall s, \forall t \hspace{0.8cm}(P_{s,t}^{i})\label{dual_solar} \\
& \sum\limits_{i\in B_{t}}\lambda2_{t}^i- \sum\limits_{i\in B_{f}}\lambda2_{t}^i - \underline{\mu5}_t^l+ \overline{\mu5}_t^l + \underline{\mu6}_t^l- \overline{\mu6}_t^l=0 \nonumber \\
& \hspace{5.8cm}\forall l, \forall t  \hspace{0.6cm} (P_t^{l})\label{dual_1c} \\
& \lambda1_t^g - \overline{\mu1}^{g, seg}_t \leq c_g^{seg} \hspace{1.4cm}\forall g, \forall t, \forall seg \hspace{0.5cm} (P_t^{g, seg})\label{dual_1d} \\
& -\lambda2_{t}^i - \overline{\mu4}_t^{d,i} \leq -K \hspace{2.4cm}\forall i, \forall t \hspace{0.7cm}(P_t^{d,i})\label{dual_1e} \\
& \sum\limits_{l\in \mathcal{L}^{fr}_{i}} \frac{\underline{\mu5}_t^l-\overline{\mu5}_t^l}{x_l} - \sum\limits_{l\in \mathcal{L}^{to}_{i}} \frac{\underline{\mu5}_t^l-\overline{\mu5}_t^l}{x_l}=0 \hspace{0.4cm}\forall t, \forall i \hspace{0.3cm}(\theta_{i}^t)\label{dual_1f} \\
& \underline{\mu7}_{t}^{l}-\mu8_{t} \leq 0 \hspace{3.4cm}\forall l, \forall t\hspace{0.7cm}(SC_{t}^l)\\
& \sum\limits_{t} \bigl(\begin{smallmatrix} \sum\limits_{s} u_t^{s,i} + \sum\limits_{i} u_t^{D,i}\end{smallmatrix}\bigr) \leq E \label{dual_1g_E}\\
& \overline{\mu}, \underline{\mu} \geq 0, \lambda \label{dual_1g}
\end{alignat}
\end{subequations}

$E$ in \eqref{dual_1g_E} is the budget of uncertainty and is decided by the system operator.
The objective function in \eqref{dual_1a} consists of non-linear terms that are formed due to binary-to-continuous variable multiplication in the dual formulation. The binary-to-continuous terms are associated with the generation units and power lines. The dual problem \eqref{dual} with non-linear terms is solved by performing the linearization as given in \eqref{dual_linear}. The binary-to-continuous term in \eqref{dual_linear_a} is linearized in \eqref{dual_linear_b}-\eqref{dual_linear_d}. Here, $I_g^t$ is a binary variable. To linearize the expression, two non-negative continuous variables $\Phi$ and $\psi$ having $0$ as a lower bound and an arbitrary large number M as an upper bound are considered. The linearization of the remaining binary-to-continuous terms is performed using the same method.
\begin{subequations}\label{dual_linear}
\begin{alignat}{2}
& \Phi_g^t = I_g^t\underline{\mu}_{g}^{t},\hspace{0.4cm} I_g^t \in \begin{Bmatrix}0,1\end{Bmatrix} \label{dual_linear_a} \\
& \Phi_g^t = \underline{\mu}_{g}^{t} - \psi_g^t \label{dual_linear_b} \\
& 0 \leq \Phi_g^t \leq M\cdot I_g^t \label{dual_linear_c} \\
& 0 \leq \psi_g^t \leq M\cdot (1-I_g^t) \label{dual_linear_d}
\end{alignat}
\end{subequations}
Once the condition in step (e) of C$\&$CGA is not satisfied, the solution of the dual problem is applied to add additional constraints \eqref{ab_1}-\eqref{ab_2} to the first stage problem. Using the solution of dual formulation \eqref{dual} and equations \eqref{change}, a new realization of the uncertainties is acquired by the addition of \eqref{ab_1}-\eqref{ab_2} constraints to the first stage problem. Where $\triangle$ in \eqref{change} denotes the deviation of scheduled demand and solar availability ($5\%, 10\%, 15\%$, and $20\%$).
\begin{subequations}\label{change}
\begin{alignat}{2}
& P_t^{D,i} = P_t^{D,i,0} +  \triangle P_t^{D,i}\cdot u_t^{D,i}  - \triangle P_t^{D,i}\cdot v_t^{D,i}\label{change_1} \\
& \overline{P}_t^i = P_{s,t}^{i,0} + \triangle \overline{P}_{s,t}^{i}\cdot u_t^{s,i} - \triangle \overline{P}_{s,t}^{i}\cdot v_t^{s,i}\label{change_2} 
\end{alignat}
\end{subequations}

\section{Case Study}\label{case_study}
In this section, two case studies are presented to demonstrate the competence of the proposed two-stage robust optimization problem to ensure the resiliency of power system network under the risk of wildfire. The first case study uses 6-bus system while the second one uses IEEE 118-bus system. The proposed method is implemented on a personal computer having 3.60 GHz processor using CPLEX 6.5.
\subsection{6-Bus Power System}
In this section, a 6-bus system is considered under the risk of wildfire ignition. It is composed of 3 generation units and 7 power lines. The characteristics of generation units and power lines are given in Tables \ref{generation_unit} and \ref{transmission_line} respectively. The demand of electricity on buses 3, 4, and 5 is 20$\%$, 40$\%$, and 40$\%$ of total generation capacity respectively. The generation units $\text{G}_1$, $\text{G}_2$, and $\text{G}_3$ are connected to buses 1, 2, and 3 respectively. The cost curve of generation units is piecewise with 3 segments.

\begin{table}[h!]
	\vspace{-0.31cm}
	\small \centering
	\caption{\color{black}Generation Unit Characteristics} 
	\vspace{-0.2cm}
	\begin{tabular}{M{0.6cm} M{0.8cm} M{0.8cm} M{0.6cm} M{0.8cm} M{1.0cm}} \hline\hline 
		Unit& P$_{\text{min}}$ (MW) & P$_{\text{max}}$ (MW)&  \textbf{a} $(\$)$&\textbf{b} $(\$$/MW) & c $(\$$/MW$^2$)   \\\hline
        G$_1$  & 100& 220 &177 &13.5 & 0.00045\\
	    G$_2$  &10 &100 &130 &40 & 0.001\\
	    G$_3$  &10 &40  & 137&17.7 &0.005\\\hline\hline
	\end{tabular}
	\vspace{-0.21cm}
	\label{generation_unit}
\end{table}

\begin{table}[h!]
	\vspace{-0.31cm}
	\small \centering
	\caption{\color{black}Transmission Line Characteristics} 
	\vspace{-0.2cm}
	\begin{tabular}{M{0.6cm} M{0.8cm} M{0.8cm} M{1.2cm} M{2.4cm}} \hline \hline 
		ID& From Bus & To Bus & Impedance (p.u.)& Maximum Rating (MW) \\ \hline 
        L$_1$  & 1 & 2  &0.170 &200\\
	    L$_2$  &2 &3 &0.037 &100\\
	    L$_3$  &1 &4 &0.258 & 100\\
	    L$_4$  & 2& 4  & 0.197& 100\\
	    L$_5$  &4 &5 & 0.037& 100\\
	    L$_6$  &5 &6 &0.140 &100\\
	    L$_7$  &3 &6 &0.018 &100\\\hline\hline 
	\end{tabular}
	\vspace{-0.21cm}
	\label{transmission_line}
\end{table}
\subsubsection{Status of transmission lines at different risk tolerances}
In this section, the impact of various risk tolerances on the energization/de-energization of lines is assessed. Risk tolerance is defined by the system operator and informs how much risk operator is willing to take to energize the power lines under wildfire ignition scenario.

\textit{Scenario 1 - (Status of transmission lines at 0 risk tolerance)}:
In the first scenario, the status of lines during 24-hour period under 0 risk tolerance is shown in Table \ref{risk_solar_24_period}. 
0 risk tolerance indicates that the network does not withstand the fire ignition score and lines are de-energized as soon as score exceeds 0. In this case, lines are de-energized for large number of hours and it leads to high objective \$1.290M. 
Lines $L_6$ and $L_7$ are vulnerable because fire ignition score exists on these lines and it leads to de-energization for 24-hour period. The network is protected against wildfire ignition in this scenario but less lines are energized and thereby renders high objective value.
\begin{table}[h!]
	\vspace{-0.31cm}
	\small \centering
	\caption{\color{black}Status of Lines at 0 Risk Tolerance during 24-Hour Period} 
	\vspace{-0.2cm}
	\begin{tabular}{cc} \hline \hline
		Lines& Hours (1-24)\\ \hline 
       $L_1$  &0 0 0 0 1 1 0 0 0 0 0 0 0 0 0 0 0 0 0 0 0 0 0 0\\
	    $L_2$ &0 0 0 0 1 1 0 0 0 0 0 0 1 0 0 0 1 1 0 0 0 0 0 1\\
	    $L_3$ &1 1 1 1 1 1 0 0 0 0 0 0 1 1 0 0 1 1 1 0 0 0 0 1\\
	    $L_4$ &0 0 0 0 1 1 0 0 0 0 0 0 0 0 0 0 0 0 0 0 0 0 0 0\\
	    $L_5$ &1 1 1 1 1 1 0 0 0 0 0 0 1 1 0 0 1 1 1 0 0 0 0 1\\
	    $L_6$ &0 0 0 0 0 0 0 0 0 0 0 0 0 0 0 0 0 0 0 0 0 0 0 0\\
	    $L_7$ &0 0 0 0 0 0 0 0 0 0 0 0 0 0 0 0 0 0 0 0 0 0 0 0\\\hline \hline
	\end{tabular}
	\vspace{-0.21cm}
	\label{risk_solar_24_period}
\end{table}

\textit{Scenario 2 - (Status of transmission lines at 0.5 risk tolerance)}:
In the second scenario, the risk tolerance of 0.5 is considered and it renders the status of lines as shown in Table \ref{risk_solar_24_period_1}. The lines remain energized if the fire ignition score is less than 0.5 otherwise the lines are de-energized. As the risk tolerance is high as compare to Scenario 1, more lines are energized, and the objective reduces to \$52,941. The highlighted values by magenta color show the change from Scenario 1.
Lines $L_6$ and $L_7$ are de-energized during 0 risk tolerance but energized for 13 hours during 0.5 risk tolerance. 
This network renders operation risk but more lines are energized and it reduces the objective value.
\begin{table}[h!]
	\vspace{-0.31cm}
	\small \centering
	\caption{\color{black}Status of Lines at 0.5 Risk Tolerance during 24-Hour Period} 
	\vspace{-0.2cm}
	\begin{tabular}{cc} \hline \hline
		Lines& Hours (1-24)\\  \hline
       $L_1$  &\textcolor{magenta}{1} \textcolor{magenta}{1} \textcolor{magenta}{1} \textcolor{magenta}{1} 1 1 0 0 0 0 0 0 \textcolor{magenta}{1} \textcolor{magenta}{1} \textcolor{magenta}{1} 0 \textcolor{magenta}{1} \textcolor{magenta}{1} \textcolor{magenta}{1} \textcolor{magenta}{1} 0 0 0 \textcolor{magenta}{1}\\
	    $L_2$ &\textcolor{magenta}{1} \textcolor{magenta}{1} \textcolor{magenta}{1} \textcolor{magenta}{1} 1 1 \textcolor{magenta}{1} \textcolor{magenta}{1} \textcolor{magenta}{1} \textcolor{magenta}{1} \textcolor{magenta}{1} \textcolor{magenta}{1} 1 \textcolor{magenta}{1} \textcolor{magenta}{1} \textcolor{magenta}{1} 1 \textcolor{magenta}{0} \textcolor{magenta}{1} \textcolor{magenta}{1} \textcolor{magenta}{1} \textcolor{magenta}{1} \textcolor{magenta}{1} 1\\
	    $L_3$ &1 1 1 1 1 1 \textcolor{magenta}{1} \textcolor{magenta}{1} \textcolor{magenta}{1} \textcolor{magenta}{1} \textcolor{magenta}{1} \textcolor{magenta}{1} 1 1 \textcolor{magenta}{1} \textcolor{magenta}{1} 1 1 1 \textcolor{magenta}{1} \textcolor{magenta}{1} \textcolor{magenta}{1} \textcolor{magenta}{1} 1\\
	    $L_4$ &\textcolor{magenta}{1} \textcolor{magenta}{1} 0 \textcolor{magenta}{1} 1 1 \textcolor{magenta}{1} \textcolor{magenta}{1} \textcolor{magenta}{1} \textcolor{magenta}{1} \textcolor{magenta}{1} \textcolor{magenta}{1} \textcolor{magenta}{1} \textcolor{magenta}{1} \textcolor{magenta}{1} \textcolor{magenta}{1} 0 \textcolor{magenta}{1} \textcolor{magenta}{1} \textcolor{magenta}{1} \textcolor{magenta}{1} \textcolor{magenta}{1} \textcolor{magenta}{1} \textcolor{magenta}{1}\\
	    $L_5$ &1 1 1 1 1 1 \textcolor{magenta}{1} \textcolor{magenta}{1} \textcolor{magenta}{1} \textcolor{magenta}{1} \textcolor{magenta}{1} \textcolor{magenta}{1} 1 1 \textcolor{magenta}{1} \textcolor{magenta}{1} \textcolor{magenta}{0} 1 1 \textcolor{magenta}{1} \textcolor{magenta}{1} \textcolor{magenta}{1} \textcolor{magenta}{1} 1\\
	    $L_6$ &0 0 \textcolor{magenta}{1} 0 0 0 0 0 0 0 \textcolor{magenta}{1} 0 \textcolor{magenta}{1} \textcolor{magenta}{1} 0 0 \textcolor{magenta}{1} \textcolor{magenta}{1} 0 0 0 0 0 0\\
	    $L_7$ &0 0 \textcolor{magenta}{1} \textcolor{magenta}{1} 0 0 0 0 0 0 0 0 \textcolor{magenta}{1} \textcolor{magenta}{1} 0 \textcolor{magenta}{1} \textcolor{magenta}{1} \textcolor{magenta}{1} 0 0 0 0 0 0\\\hline \hline 
	\end{tabular}
	\vspace{-0.21cm}
	\label{risk_solar_24_period_1}
\end{table}

\textit{Scenario 3 - (Status of transmission lines at 1.0 risk tolerance)}:
In this scenario, the risk tolerance of 1.0 is considered and it leads to the energization of all lines during 24-hour period. 
The high risk tolerance poses high wildfire ignition risk. It leads to all lines energization during whole day and the objective reduces to \$607 which is lower than Scenario 2.

\subsubsection{Transmission lines risk of operation quantification}
In this section, the conservative and cumulative fire risk intakes for the network operation are assessed. The conservative case considers sum of fire ignition scores for all lines during 1-hour period and it leads to less risk, while the cumulative case considers sum of fire ignition score for all lines during 24-hour period and it leads to high risk. In each case, different risk tolerance levels are considered to find the appropriate operation scenario.

\textit{Scenario 1 - (Risk of operation quantification with conservative fire risk intake)}:
The conductor clashing score for energized power lines with conservative fire risk intake under different risk tolerance levels for 24-hour period is shown in Table \ref{risk_solar_24_period_4}. In Level 1, all lines have 0 risk of operation because the risk tolerance is 0. In Level 2, the risk of operation increases as the risk tolerance is increased from 0 to 0.5. 
The Level 3 has highest risk of operation due to high risk tolerance of 1.0. Line $L_6$ is the most dangerous line with a risk of operation 3.74. The 0 risk of operation in line $L_5$ indicates that it has no threat for operation. Lines $L_2$, $L_3$, and $L_5$ are crucial for the network because the operator can rely on these lines to serve the demand.

\begin{table}[h!]
	\vspace{-0.31cm}
	\small \centering
	\caption{\color{black}Quantification of Risk of Operation With Conservative Fire Risk Intake For 24-Hour Period } 
	\vspace{-0.2cm}
	\begin{tabular}{cccccccc} \hline \hline 
		& $L_1$ & $L_2$ & $L_3$ & $L_4$ & $L_5$ & $L_6$ & $L_7$\\ \hline 
        Level 1  &0 & 0 & 0 & 0 & 0 & 0 & 0\\
	    Level 2 & 1.22& 1.41 & 1.38 & 2.62 & 0 &  1.17&0.63 \\
	    Level 3 &3.07 &1.41  & 1.38 & 2.8 & 0 &3.74  &3.14 \\\hline \hline 
	\end{tabular}
	\vspace{-0.21cm}
	\label{risk_solar_24_period_4}
\end{table}

\textit{Scenario 2 - (Risk of operation quantification with cumulative fire risk intake)}:
The quantification of the risk of operation with cumulative fire risk intake for 24-hour period is shown in Table \ref{risk_solar_24_period_5}. The magenta color highlighted values show the change from Table \ref{risk_solar_24_period_4}. In Level 1, the objective is increased to \$1.301M as compared to conservative fire risk intake.
In Level 2, the objective is \$1.042M at risk tolerance of 0.5 and the decrease in objective occurs because high risk tolerance leads to more lines energization. In Level 3, the objective reduces to \$840,978 at risk tolerance of 1.0. 
The objective for cumulative fire risk intake is higher than the conservative risk intake.
The lines $L_1$, $L_5$, $L_6$, and $L_7$ have zero risk of operation and these are safest to energize. Line $L_2$ is dangerous while the lines $L_3$ and $L_4$ are the most dangerous to operate.
\begin{table}[h!]
	\vspace{-0.31cm}
	\small \centering
	\caption{\color{black}Quantification of Risk of Operation With Cumulative Fire Risk Intake For 24-Hour Period} 
	\vspace{-0.2cm}
	\begin{tabular}{cccccccc} \hline \hline 
		& $L_1$ & $L_2$ & $L_3$ & $L_4$ & $L_5$ & $L_6$ & $L_7$\\ \hline
        Level 1  &0 & 0 & 0 & 0 & 0 & 0 & 0\\
	    Level 2 & \textcolor{magenta}{0}& \textcolor{magenta}{0.04} & \textcolor{magenta}{0.32} & \textcolor{magenta}{0.14} & 0 & \textcolor{magenta}{0}&\textcolor{magenta}{0} \\
	    Level 3 &\textcolor{magenta}{0} &\textcolor{magenta}{0.06} & \textcolor{magenta}{0.48} & \textcolor{magenta}{0.46} & 0 &\textcolor{magenta}{0}  &\textcolor{magenta}{0} \\\hline \hline 
	\end{tabular}
	\vspace{-0.21cm}
	\label{risk_solar_24_period_5}
\end{table}

\subsubsection{Consequence of different risk tolerances on the operation cost by varying budget of uncertainty}
\begin{figure}[bht] 
    \centering
       \includegraphics[width=0.45\textwidth]{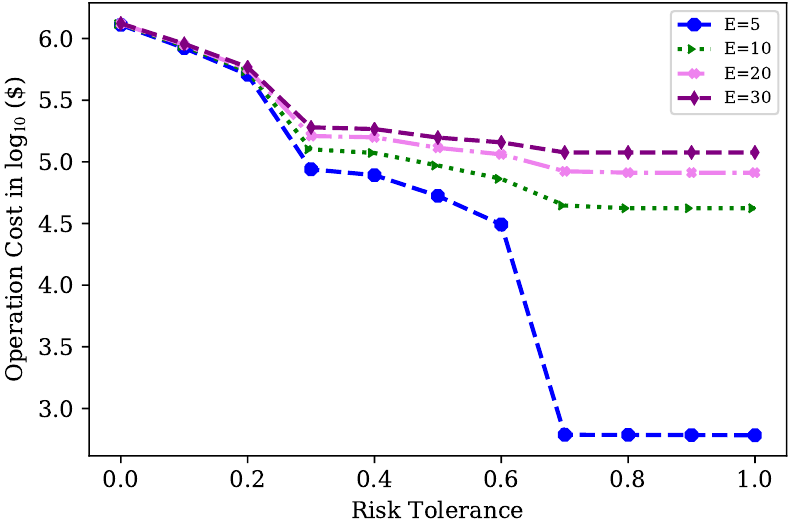}
    \caption{Impact of risk tolerance on the operation cost under different budget of uncertainties in 6-bus test case}
    \label{gen_cost_tolerance_dev} 
\end{figure}

In this section, the change in objective by varying risk tolerance level under different budget of uncertainties is assessed. Here, the generation unit $\text{G}_3$ is replaced with 100MW solar generation. Fig. \ref{gen_cost_tolerance_dev} shows the changes in operation cost with respect to risk tolerance when budget of uncertainty is varied ($E$=5, 10, 20, 30) and deviation in load demand and solar generation is 10\%. The objective is in log base 10 due to wide variation. The increase in risk tolerance decreases the operation cost. At lower budget of uncertainty $E$=5, the operation cost reduces to \$606 at 0.7 risk tolerance. A sharp increase in operation cost at 0.7 tolerance exists because the demand is higher. The higher changes in demand cause the increase in operation cost. The increase in budget of uncertainty from 5 to 10 leads to the increase in minimum operation cost to \$42,021 at 0.7 risk tolerance. To operate the network at the lowest cost and under the uncertainty scenario, it is recommended to have risk tolerance 0.7 and budget of uncertainty 5.

\subsubsection{Impact of budget of uncertainty on operation cost by varying the uncertainty in demand and solar generation}
\begin{figure}[bht] 
    \centering
       \includegraphics[width=0.45\textwidth]{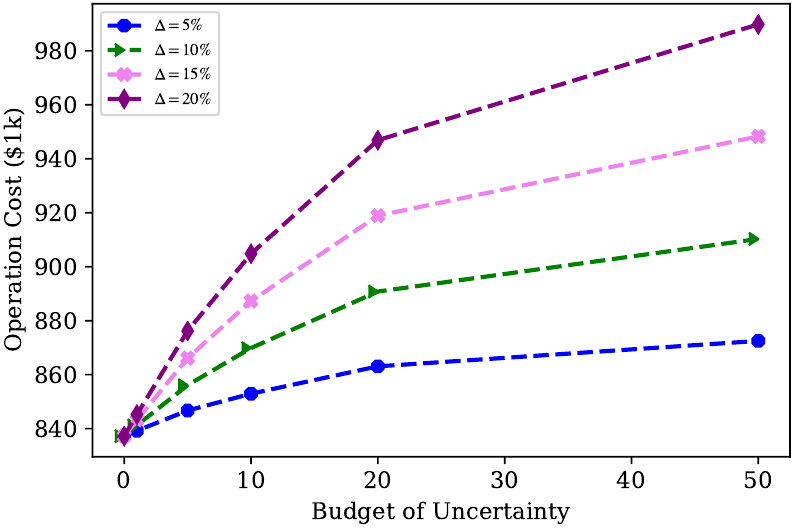}
    \caption{Change in operation cost based on budget of uncertainty when $\mathcal{E}=0.1$ and different deviations are considered in 6-bus test case}
    \label{gen_cost_tolerance_dev_1} 
\end{figure}

In this section, the relationship between the budget of uncertainty and the operation cost is assessed. The increase in budget of uncertainty leads to the increase in the operation cost as shown in Fig. \ref{gen_cost_tolerance_dev_1}. The graph shows the relationship between these variables based on different deviations in demand and solar generation ($\Delta$=5\%, 10\%, 15\%, and 20\%). At 5\% deviation, the minimum operation cost of \$837,220 exists at budget uncertainty 0 and the maximum \$872,549 exists at budget of uncertainty 50. By increasing the deviation from 5\% to 20\%, the operation cost increases proportionally. It results in direct relationship between the deviation and the operation cost. 

\subsubsection{Assessing the impact of solar integration on the operation cost}
\begin{figure}[bht] 
    \centering
       \includegraphics[width=0.45\textwidth]{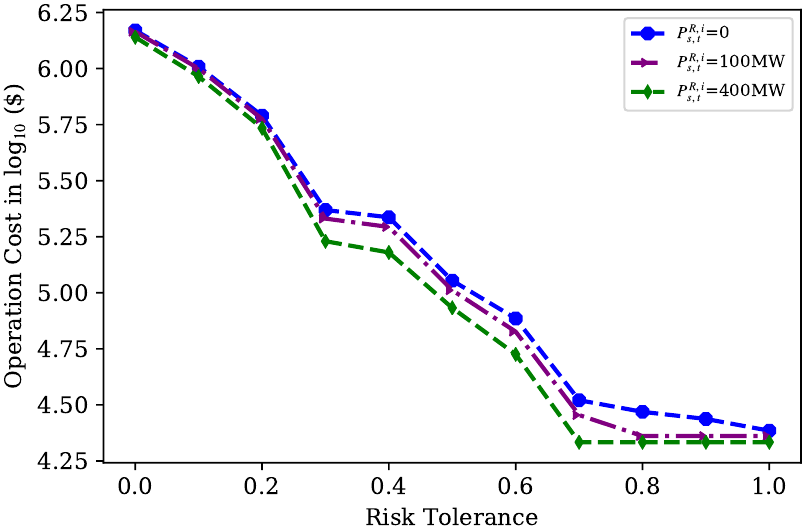}
    \caption{Impact of different levels of solar penetration on the operation cost in 6-bus test case}
    \label{gen_cost_tolerance_dev_2} 
\end{figure}

The impact of solar integration on different buses to reduce the operation cost is shown in Fig. \ref{gen_cost_tolerance_dev_2}. generation units $\text{G}_1$ and $\text{G}_2$ are connected to buses 1 and 2 respectively. The solar integration ${P}_{1,t}^{R,3}$, ${P}_{2,t}^{R,4}$, ${P}_{3,t}^{R,5}$, and ${P}_{4,t}^{R,6}$ occurs on buses 3, 4, 5, and 6 respectively in the 6-bus test case where $t$ ranges 1-24 hours. The operation cost on y-axis is in log$_{\text{10}}$ due to large range of cost with different risk tolerance levels. With no solar integration, the maximum operation cost at risk tolerance 0 is \$1.482M and it reduces to \$24,259 at risk tolerance 1. When 25MW solar is integrated, the maximum operation cost reduces to \$1.463M (0 risk tolerance) and the minimum cost reduces to \$22,963 (1 risk tolerance). By further increasing the solar integration to 100MW on each bus, the maximum operation cost reduces to \$1.381M and the minimum cost reduces to \$21,562. The increase in solar integration leads to the decrease in operation cost, it makes the network more resilient, and thus mitigates the impact of wildfire.
\subsubsection{Comparison of resilient operation under centralized and distributed solar integration by considering  uncertainty}
\begin{figure}[bht] 
    \centering
       \includegraphics[width=0.45\textwidth]{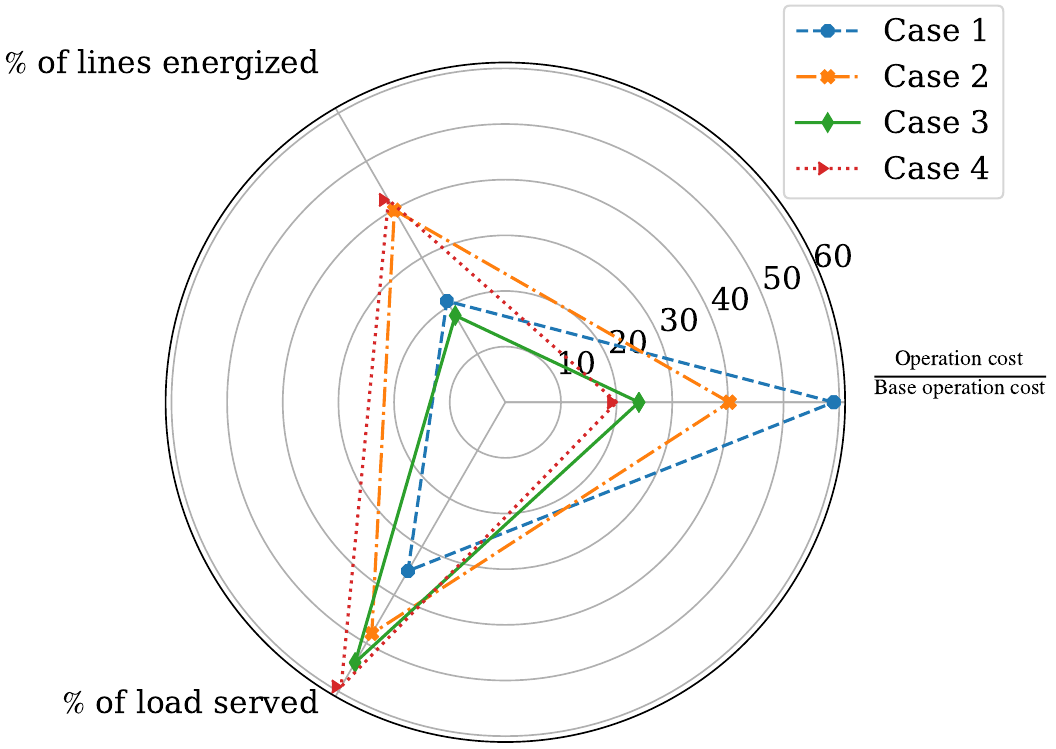}
    \caption{Effect of centralized and distributed solar generations under uncertainty on operation cost, \% of lines energized, and load served}
    \label{spider_chart} 
\end{figure}

A comparison chart that demonstrates the impact of distributed and centralized solar integration on the operation cost, percentage of lines energized, and percentage of load served under uncertainty in demand and solar is shown in Fig. \ref{spider_chart}. The percentage of load served is calculated by dividing the demand served at bus $i$ and time $t$ to the scheduled demand and the percentage of lines energized is determined by dividing number of lines energized to the total lines in the network. Here, the budget of uncertainty is 5, risk tolerance is 0.1, and it results in operation cost nominal value \$21,165.

\textit{Case 1 - (Centralized solar integration with cumulative fire risk intake)}:
In this case, the centralized solar is integrated in the 6-bus test case during 24-hour period under wildfire ignition risk and uncertainty conditions. The solar generation of 48MW is integrated on bus 3 only.  This case results in operation cost of \$1.24M and it is 59 times the nominal value. Here, the load served is 35\% and 21\% lines are energized. 

\textit{Case 2 - (Centralized solar integration with conservative fire risk intake)}: By considering conservative fire risk intake and centralized solar integration of 48MW on bus 3, the operation cost is \$851,488 and it is 31\% less than Case 1. Here, more 40\% of the lines are energized and the percentage increase in energized lines from Case 1 is 91\%. The load served is 48\% that is greater than Case 1.

\textit{Case 3 - (Distributed solar integration with cumulative fire risk intake)}: In this case, the solar generation of 48MW is equally distributed on buses 3, 4, 5, and 6. The operation cost reduces to \$500,339 as compared to \$1.24M in Case 1. As compared to Case 1, the operation cost is decreased by 60\%, the energized power lines are reduced by 14\%, and the load served is increased from 35\% to 54\%.

\textit{Case 4 - (Distributed solar integration with conservative fire risk intake)}: By considering conservative fire risk intake and solar generation of 48MW equally distributed on buses 3, 4, 5, and 6, the operation cost is \$414,965. In comparison to Case 3, the operation cost is decreased by 17\%, the energized lines are increased by 133\%, and load served is increased from 54\% to 59\%. As compared to Case 2, the operation cost is reduced by 51\%, the increased in energized lines is 4\%, and the demand served is increased from 48\% to 59\%.

In result, centralized solar generation with cumulative fire risk intake has very high operational cost. Distributed solar generation with cumulative risk intake is better because with few number of energized lines, the operation cost is decreased significantly, it has less operating risk, and more load is served. Although, distributed solar generation with conservative fire risk intake has lower operation cost as compared to solar generation with cumulative fire risk, but it has high number of lines energized. Being more liberal is better in terms of operation cost but more lines are energized so the risk of fire ignition is more. Less number of lines energized, lower operation cost, and higher value of load served is the most appropriate scenario and it occurs in case of distributed solar generation with cumulative risk intake. Distributed solar is helping us to have less lines energized, serve more customers and have less operation cost. Thus, if we have renewable generation in a distributed manner, it is going to help us mitigate the risk of fire ignition and serve more customers. 
\subsection{IEEE 118-Bus Power System}
\begin{figure}[h!] 
    \centering
       \includegraphics[width=0.45\textwidth]{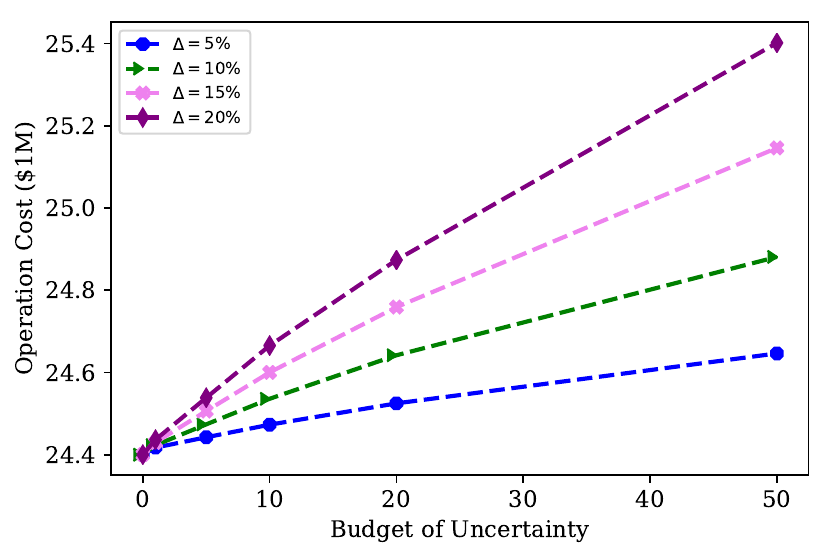}
    \caption{Impact of budget of uncertainty on the operation cost when $\epsilon$=0.1 and deviation in demand and solar generation are considered for IEEE 118-bus system}
    \label{bus_118} 
\end{figure}

In this section, the risk-averse resilient operation of IEEE 118-bus system under fire ignition risk is considered.
The impact of budget of uncertainty on the operation cost is assessed for different deviations in demand and solar generation as shown in Fig. \ref{bus_118}. 
At 5\% deviation, the minimum operation cost at 0 budget of uncertainty is \$24.40M and the maximum operation cost increases to \$24.64M at 50 budget of uncertainty. By varying the deviation from 5\% to 20\%, the operation cost rises and it indicates that the deviation from nominal value leads to higher operation cost under the same risk tolerance. At 20\% deviation the minimum operation cost is \$24.40M but the maximum operation cost increases to \$25.40M.
With the increase in budget of uncertainty, the operation cost increases. By increasing the deviation, the operation cost is increasing even more. For example, with budget of uncertainty 50 and deviation 20\%, the operation cost is 1.04 times the cost at budget of uncertainty 5 and deviation 5\%.

\section{Conclusions}\label{conclusions}
This paper presented a two-stage robust optimization problem to ensure the risk-averse resilient operation of electricity grid under the risk of fire hazard weather conditions. The power lines are quantified based on the risk of fire hazard score and resilient operation of power system is ensured to supply power to customers during such conditions. The robust optimization problem finds a balance between de-energization of lines and the customers served. Different levels of penetration of solar generation to mitigate the impact of fire hazard situations on the energization of customers is assessed. The impact of centralized and distributed solar generation based on cumulative and conservative fire risk intakes to mitigate the risk of fire hazard is evaluated. System operator can serve the load and protect the system from wildfire hazard simultaneously using the presented optimization problem. The case studies of 6- and IEEE 118-bus systems validate the presented risk-averse resilient operation of power system using the robust optimization problem.
\ifCLASSOPTIONcaptionsoff
\fi
\bibliographystyle{IEEEtran}

\bibliography{references}

\end{document}